\documentclass[sigconf,natbib=true,anonymous=false]{acmart}

\usepackage{graphicx}
\usepackage{amsmath}
\usepackage{booktabs}
\usepackage{algorithm}
\usepackage{algorithmic}
\usepackage{amsfonts}
\usepackage{multirow}[c]
\usepackage{makecell}
\usepackage{subfigure}
\usepackage{color}
\usepackage{bm}
\usepackage{epstopdf}
\usepackage{url}
\usepackage[cal=cm]{mathalfa}
\usepackage{balance}
\usepackage{threeparttable}
\usepackage{wrapfig}
\usepackage{tabularx}
\usepackage{array}
\usepackage{enumitem}
\usepackage{appendix}

\setlist[itemize]{leftmargin=*}

\makeatletter
\newcommand\notsotiny{\@setfontsize\notsotiny\@vipt\@viipt}
\makeatother

\AtBeginDocument{%
  \providecommand\BibTeX{{%
    \normalfont B\kern-0.5em{\scshape i\kern-0.25em b}\kern-0.8em\TeX}}}

\author{Xiaodong Li}
\affiliation{
  \institution{Institute of Information Engineering, Chinese Academy of Sciences \\School of Cyber Security, UCAS$^\dag$}
  \city{Beijing, China} \\
  \country{lixiaodong@iie.ac.cn}
 }

\author{Hengzhu Tang}
\affiliation{
  \institution{Baidu Inc}
  \city{Beijing, China} \\
  \country{hengzhutang@gmail.com}
 }

\author{Jiawei Sheng}
\affiliation{
  \institution{Institute of Information Engineering, Chinese Academy of Sciences}
  \city{Beijing, China} \\
  \country{shengjiawei@iie.ac.cn}
 }
 \authornote{Corresponding author.\\ $^\dag$University of Chinese Academy of Sciences.}

\author{Xinghua Zhang}
\affiliation{
  \institution{Institute of Information Engineering, Chinese Academy of Sciences \\ School of Cyber Security, UCAS}
  \city{Beijing, China} \\
  \country{zhangxinghua@iie.ac.cn}
 }

\author{Li Gao}
\affiliation{
  \institution{Baidu Inc}
  \city{Beijing, China} \\
  \country{gaoli.sinh@gmail.com}
 }
 \authornotemark[1]

\author{Suqi Cheng}
\affiliation{
  \institution{Baidu Inc}
  \city{Beijing, China} \\
  \country{chengsuqi@gmail.com}
 }

\author{Dawei Yin}
\affiliation{
  \institution{Baidu Inc}
  \city{Beijing, China} \\
  \country{yindawei@acm.org}
 }

\author{Tingwen Liu}
\affiliation{
  \institution{Institute of Information Engineering, Chinese Academy of Sciences \\ School of Cyber Security, UCAS}
  \city{Beijing, China} \\
  \country{liutingwen@iie.ac.cn}
 }

\copyrightyear{2025}
\acmYear{2025}
\setcopyright{rightsretained}
\acmConference[KDD '25]{Proceedings of the 31st ACM SIGKDD Conference on Knowledge Discovery and Data Mining V.1}{August 3--7, 2025}{Toronto, ON, Canada}
\acmBooktitle{Proceedings of the 31st ACM SIGKDD Conference on Knowledge Discovery and Data Mining V.1 (KDD '25), August 3--7, 2025, Toronto, ON, Canada}
\acmDOI{10.1145/3690624.3709220}
\acmISBN{979-8-4007-1245-6/25/08}

\begin{document}
\title{Exploring Preference-Guided Diffusion Model for\\ Cross-Domain Recommendation}

\renewcommand{\shorttitle}{Exploring Preference-Guided Diffusion Model for Cross-Domain Recommendation}
\renewcommand{\shortauthors}{Xiaodong Li et al.}

\begin{abstract}
Cross-domain recommendation (CDR) has been proven as a promising way to alleviate the cold-start issue, in which the most critical problem is how to draw an informative user representation in the target domain via the transfer of user preference existing in the source domain. 
Prior efforts mostly follow the embedding-and-mapping paradigm, which first integrate the preference into user representation in the source domain, and then perform a mapping function on this representation to the target domain. However, they focus on mapping features across domains, neglecting to explicitly model the preference integration process, which may lead to learning coarse user representation.
Diffusion models (DMs), which contribute to more accurate user/item representations due to their explicit information injection capability, have achieved promising performance in recommendation systems.
Nevertheless, these DMs-based methods cannot directly account for valuable user preference in other domains, leading to challenges in adapting to the transfer of preference for cold-start users. Consequently, the feasibility of DMs for CDR remains underexplored.
To this end, we explore to utilize the explicit information injection capability of DMs for user preference integration and propose a \textit{Preference-Guided Diffusion Model} for CDR to cold-start users, termed as \textbf{DMCDR}. Specifically, we leverage a preference encoder to establish the preference guidance signal with the user's interaction history in the source domain.
Then, we explicitly inject the preference guidance signal into the user representation step by step to guide the reverse process, and ultimately generate the personalized user representation in the target domain,
thus achieving the transfer of user preference across domains. Furthermore, we comprehensively explore the impact of six DMs-based variants on CDR. Extensive experiments on three real-world CDR scenarios demonstrate the superiority of our DMCDR over SOTA methods and six DMs-based variants.

\end{abstract}

\ccsdesc[500]{Information systems~Recommender systems}

\keywords{Cross-Domain Recommendation, Cold-Start Recommendation, \\ Diffusion Models}

\maketitle

\begin{figure}[t!]
\begin{center}
\includegraphics[width=8.0cm]{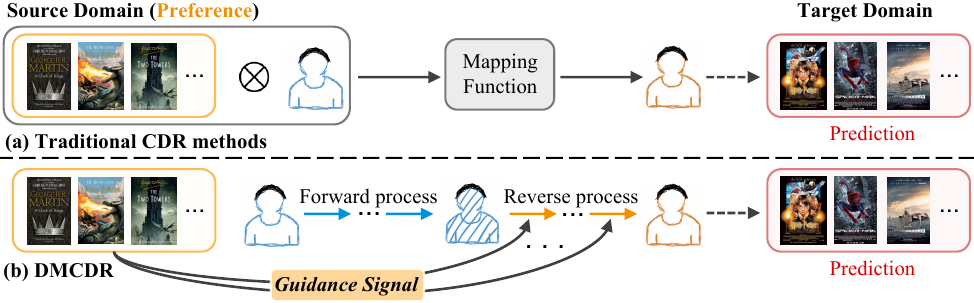}
\caption{An illustration of (a) traditional CDR methods, where $\otimes$ denotes the integration operation; (b) our DMCDR.}
\label{motivation}
\end{center}
\end{figure}

\section{Introduction} 

Recommendation systems (RS)~\cite{ncf,mmnr,pkef} have long been an important topic in many online platforms, such as Amazon (e-commerce), TikTok (online video) and Facebook (social media). 
With the advent of the digital age, users prefer to participate in multiple domains (platforms) with different purposes, \textit{e.g.}, buying books on Amazon and watching short videos on TikTok. Therefore, how to make recommendations to new coming (cold-start) users across domains has become a key issue in RS. 
To mitigate the \textit{cold-start issue}~\cite{cdrnp,mlhin,mvdg}, \textbf{\textit{cross-domain recommendation (CDR)}}~\cite{emcdr,ptupcdr,cdrnp,ccdr} has attracted significant attention in recent years, which leverages user's interaction history from an informative source domain to improve the recommendation performance in the target domain.

The core task of CDR is to draw an informative user representation in the target domain via the transfer of user preference existing in the source domain.
To achieve this, CDR works~\cite{emcdr,sscdr,ptupcdr,remit,cdrnp} mostly adhere to the embedding-and-mapping paradigm (as shown in Figure~\ref{motivation}(a)), where they first incorporate the preference into user representation in the source domain, and then perform a mapping function on this representation to the target domain.
For example, PTUPCDR~\cite{ptupcdr} draws upon the vanilla matrix multiplication operation to incorporate preference into user representation in the source domain, and precisely adapts MAML~\cite{mamlfa} to map the representation.
CDRNP~\cite{cdrnp} carries out the similar preference integration, and then models preference correlations among users based on Neural Process~\cite{np} for more accurate feature mapping.
Despite achieving promising results, their emphasis lies on mapping features across domains, and the modeling of preference integration process is overlooked, which may lead to learning coarse user representation.

Recently, \textit{Diffusion Models} (DMs)~\cite{ddpm,dulunt} have emerged promising results in the field of image synthesis~\cite{dmbgans} and text generation~\cite{diffusion-lm}, which gradually add noise to the input sample and then iteratively learn the reverse reconstruction process.
Currently, several studies~\cite{ddpm,dreamrec,mcdrec,difashion} have demonstrated the explicit \textit{information injection} capability of DMs to generate more accurate user/item representations in RS.
For instance, MCDRec~\cite{mcdrec} incorporates the multi-modal knowledge into item representation modeling via DMs.
Likewise, DreamRec~\cite{dreamrec} applies DMs to generate the oracle items with user's sequential behaviors. 
However, it is worth noting that these DMs-based methods cannot directly account for valuable user preference in other domains, leading to challenges in adapting to the transfer of preference for cold-start users.
Consequently, the feasibility and great potential of DMs for CDR is mostly underexplored.
In addition, direct adaptation of existing DMs-based recommenders to transfer user preference across domains also brings sub-optimal performance, which has been verified in Section~\ref{variant_study}.

To bridge this research gap, we explore to take advantage of the explicit information injection capability of DMs for user preference integration
and propose a simple yet effective approach \textbf{DMCDR} (as shown in Figure~\ref{motivation}(b)), which applies the \textit{Preference-Guided Diffusion Model} to generate the personalized representation for cold-start users in the target domain under the guidance of user preference in the source domain. Specifically, DMCDR gradually corrupts user representation by adding Gaussian noise in the forward process,  
then explicitly injects user preference into the corrupted user representation step by step to guide the reverse process and finally generate the personalized user representation in the target domain,
thereby achieving the transfer of user preference across domains. Wherein, we establish the preference guidance signal with a preference encoder on user's interaction history in the source domain, so that the generated user representation through guidance is specialised for user preference in the source domain. Taking one step further, we comprehensively explore the impact of six DMs-based variants on CDR (see Section~\ref{variant_study}), demonstrating the effectiveness and superiority of our proposed DMCDR. 

In summary, the contributions of this work are as follows:
\begin{itemize}
    \item We introduce a fresh perspective to solve the cold-start problem in CDR with DMs-based paradigm.
    \item We propose a novel Preference-Guided Diffusion Model for CDR to cold-start users, termed as DMCDR, which generates personalized user representation in the target domain under the guidance of user preference in the source domain.
    \item We comprehensively explore the impact of six DMs-based variants on CDR. And we conduct extensive experiments on three real-world CDR scenarios, demonstrating remarkable performance and superiority of DMCDR over SOTA methods (up to 32.60\% improvement) and six DMs-based variants.
\end{itemize}

\section{Preliminary}
This section introduces the background knowledge of diffusion models and defines our CDR problem settings.

\subsection{Diffusion Models}\label{DMs}
Diffusion models (DMs) have demonstrated significant performance in diverse fields including computer vision~\cite{dmbgans,ddpm,iddpm,ddim} and text generation~\cite{diffuseq,diffusion-lm}. Typically, diffusion models involve forward and reverse processes~\cite{ddpm,dulunt}, which formulates two Markov chains to model the underlying data generating distribution.

\subsubsection{\textbf{Forward process}}
Given an input sample $\bm{x}_0 \sim q(\bm{x}_0)$, the DMs gradually add Gaussian noise to $\bm{x}_0$ to construct $\bm{x}_{1:T}$ with a variance schedule $[\beta_1,\beta_2,...,\beta_T]$ in $T$ steps:
\begin{equation}
\small
\begin{split}
q(\bm{x}_{1:T}|\bm{x}_0) &= \prod_{t=1}^{T}q(\bm{x}_t|\bm{x}_{t-1}),\\
q(\bm{x}_t|\bm{x}_{t-1}) &= \mathcal{N}(\bm{x}_t;\sqrt{1-\beta_t}\bm{x}_{t-1},\beta_t\textbf{I}),
\end{split}
\label{}
\end{equation}
where $t\in\{1,\dots,T\}$ denotes the diffusion step, $\beta_t\in(0,1)$ controls the forward noise scale added at each step $t$, and $\mathcal{N}$ denotes the Gaussian distribution. As $T\to\infty$, $\bm{x}_T$ converges to a standard Gaussian noise $\bm{x}_T\sim\mathcal{N}(0,\textbf{I})$~\cite{ddpm}.

\subsubsection{\textbf{Reverse process}}
The reverse process aims to denoise $\bm{x_T}$ for $T$ steps to reconstruct the initial representation $\bm{x}_0$ using neural networks. Starting from $\bm{x}_T$, the reverse process is formulated as:
\begin{equation}
\small
\begin{split}
p_{\theta}(\bm{x}_{0:T}) &= p(\bm{x}_T)\prod_{t=1}^{T}p_{\theta}(\bm{x}_{t-1}|\bm{x}_t),\\
p_{\theta}(\bm{x}_{t-1}|\bm{x}_t) &= \mathcal{N}(\bm{x}_{t-1};\bm{\mu}_\theta(\bm{x}_t,t),\textstyle\sum_\theta(\bm{x}_t,t)),
\end{split}
\label{pre_reverse}
\end{equation}
where $\bm{\mu}_\theta(\bm{x}_t,t)$ and $\sum_\theta(\bm{x}_t,t)$ denote the mean and variance of the Gaussian distribution, which can be learned through neural networks (\textit{e.g.}, U-Net for image generation~\cite{ddpm}, Transformer for text generation~\cite{diffusion-lm}, MLP for RS~\cite{drm}) parameterized by $\theta$.

\subsubsection{\textbf{Optimization}}
The objective of training the DMs is to optimize the underlying data generating distribution $p_\theta(\bm{x}_0)$, which is equivalent to optimize the usual variational bound of negative log-likelihood:
\begin{equation}
\small
\begin{split}
\mathbb{E}[-\text{log}p_\theta(\bm{x}_0)]&\leqslant\mathbb{E}_q\left[-\text{log}p(\bm{x}_T)-\sum_{t=1}^{T}\text{log}\frac{p_\theta(\bm{x}_{t-1}|\bm{x}_t)}{q(\bm{x}_t|\bm{x}_{t-1})}\right]\\
&=\mathbb{E}_q\Bigg[\underbrace{D_{\text{KL}}(q(\bm{x}_T|\bm{x}_0)\Vert p(\bm{x}_T))}_{\mathcal{L}_T}\underbrace{-\text{log}p_\theta(\bm{x}_0|\bm{x}_1)}_{\mathcal{L}_0}\\
&\quad\quad+\sum_{t=2}^{T}\underbrace{D_{\text{KL}}(q(\bm{x}_{t-1}|\bm{x}_t,\bm{x}_0)\Vert p_\theta(\bm{x}_{t-1}|\bm{x}_t))}_{\mathcal{L}_{t-1}}\Bigg]
\end{split}
\label{pre_loss}
\end{equation}
where the first term $\mathcal{L}_T$ is a constant that does not depend on $\theta$ and thus ignorable during the optimization; the second term $\mathcal{L}_0$ represents the negative reconstruction probability of $\bm{x}_0$; and the third term $\mathcal{L}_{t-1}$ denotes minimizing the KL divergence between $q(\bm{x}_{t-1}|\bm{x}_t,\bm{x}_0)$ and $p_\theta(\bm{x}_{t-1}|\bm{x}_t)$. As suggested by~\cite{drm,diffkg}, $\mathcal{L}_0 + \sum_{t=2}^T\mathcal{L}_{t-1}$ can be simplified to $\sum_{t=1}^T\mathcal{L}_{t-1}$ for convenience. Therefore, we can optimize $\theta$ by minimizing $\sum_{t=1}^T\mathcal{L}_{t-1}$.

\subsubsection{\textbf{Inference}}\label{pre_inference}
After optimizing $\theta$, DMs randomly sample a Gaussian noise $\bm{x}_T\sim\mathcal{N}(0,\textbf{I})$ as the initial state, and leverage $p_\theta(\bm{x}_{t-1}|\bm{x}_t)$ to iteratively conduct the generation process $\bm{x}_T\to\bm{x}_{T-1}\to\dots\to\bm{x}_0$ via Eq.~(\ref{pre_reverse}).

\subsection{Problem Setting}
We use $\mathcal{D}^s$ and $\mathcal{D}^t$\footnote{We slightly abuse the notation $t$ to denote the target domain for brevity.} to denote the user-item interaction data from source domain and target domain respectively. Formally, the interaction data can be formulated as $\mathcal{D}^s=\{\mathcal{U}^s,\mathcal{V}^s,\mathcal{Y}^s\}$ and $\mathcal{D}^t=\{\mathcal{U}^t,\mathcal{V}^t,\mathcal{Y}^t\}$, where $\mathcal{U}$, $\mathcal{V}$ and $\mathcal{Y}$ denote the user set, item set and rating matrix. Here $y_{ij}\in\mathcal{Y}$ is the rating that user $u_i$ gives to item $v_j$. And the interaction history of user $u_i$ in the source domain can be denoted as $\mathcal{H}_i^s=\{v_j^s|v_j^s\in\mathcal{V}^s\}_{j=1}^{|\mathcal{H}_i^s|}$. Further, we define the users overlapped in both domains as the overlapping users $\mathcal{U}^o=\mathcal{U}^s\cap\mathcal{U}^t$, then the cold-start users that exist only in the source domain can be represented as $\mathcal{U}^{s\setminus o} = \mathcal{U}^{s}\setminus \mathcal{U}^o$. Consequently, we leverage the overlapping users $u_i^o\in\mathcal{U}^o$ to train our model. In the inference phase, given a cold-start user $u_i^{s\setminus o}\in\mathcal{U}^{s\setminus o}$, our CDR goal is to estimate the rating $\hat{y}_{ij}^t$ that user $u_i^{s\setminus o}$ would give to a candidate item $v_j^t$ in the target domain.

\begin{figure*}[t!]
\begin{center}
\includegraphics[width=15.5cm]{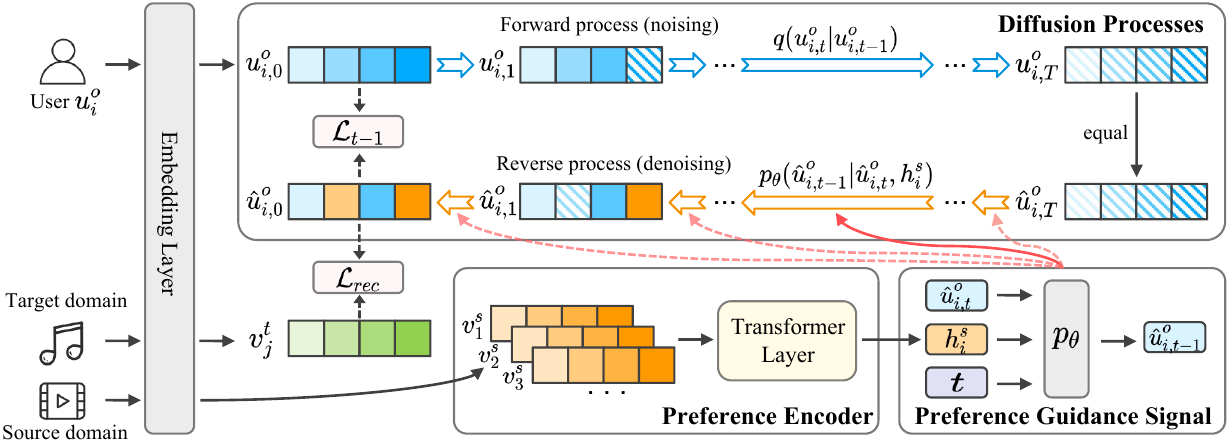}
\caption{The overall framework of DMCDR in the training phase. To achieve preference transfer, DMCDR encodes user's interaction history $\mathcal{H}_i^s$ in the source domain as preference guidance signal $\bm{h}_i^s$, then explicitly injects $\bm{h}_i^s$ into the user representation step by step to guide the reverse process, and finally generates the personalized user representation in the target domain.}
\label{main_model}
\end{center}
\end{figure*}

\section{Methodology}
This section introduces our proposed DMCDR implementing the preference-guided diffusion principle, as depicted in Figure~\ref{main_model}. We start by an embedding layer to initialize user/item embeddings as the inputs of DMCDR. Subsequently, we use a preference encoder to generate the preference guidance signal based on user's interaction history in the source domain.
Furthermore, inspired by DMs, we explain our approach to generate user representation in the target domain guided by the preference guidance signal.
To conclude, we introduce classifier-free guidance to further ensure the generated user representation align closely with preference guidance signal.

\subsection{Embedding Layer}
The embedding layer provides initialized representations for users and items in both domains. Specifically, we initialize user representation as $\bm{u}_i\in\mathbb{R}^{d_1}$, item representation as $\bm{v}_j^s\in\mathbb{R}^{d_1}$ in the source domain and $\bm{v}_j^t\in\mathbb{R}^{d_1}$ in the target domain, where $d_1$ denotes the embedding dimension.

\subsection{Preference Encoder}\label{guidance_signal}
The preference encoder aims to generate the preference guidance signal based on user's interaction history in the source domain, which is used to guide the subsequent diffusion processes. 

Intuitively, different items that users interact with in the source domain contribute differently to their preference. Therefore, we propose to encode the interaction history $\mathcal{H}_i^s$ of user $u_i$ in the source domain with a Transformer~\cite{trans} Layer:
\begin{equation}
\small
\begin{split}
\bm{c}_i^s=\text{Trans-Lay}([\bm{v}_1^s,\bm{v}_2^s,\dots,\bm{v}_{|\mathcal{H}_i^s|}^s]),
\end{split}
\label{encoder_1}
\end{equation}
where $\bm{c}_i^s\in\mathbb{R}^{|\mathcal{H}_i^s|\times{d_1}}$ is a temporary preference representation. Then, we generate the preference guidance signal $\bm{h}_i^s$ with an average pooling function for computational efficiency:
\begin{equation}
\small
\begin{split}
\bm{h}_i^s=\text{Ave-Pool}(\bm{c}_i^s),
\end{split}
\label{encoder_2}
\end{equation}
where $\bm{h}_i^s\in\mathbb{R}^{d_1}$ summarizes the personalized preference of user $u_i$ in the source domain. After that, we can leverage the preference guidance signal $\bm{h}_i^s$ to guide the subsequent diffusion processes, thus can make CDR for cold-start users in the target domain.

\subsection{Diffusion Processes} 
A simple application of DMs (described in Section~\ref{DMs}), cannot sufficiently generate the personalized user representation in the target domain, as the reverse process modelled in Eq.~(\ref{pre_reverse}) lacks the guidance from the interaction history in the source domain, resulting in the generation of non-personalized user representation. To address the above issues, we propose to guide the reverse process with user's interaction history in the source domain. Specifically, the proposed DMCDR consists of two critical processes: 1) the forward process gradually corrupts user representation by adding Gaussian noise, 2) the reverse process learns to denoise and reconstruct the user representation. During the reverse process, DMCDR considers the user's interaction history in the source domain as the preference guidance signal (described in Section~\ref{guidance_signal}), resulting in the generation of personalized user representation.

\subsubsection{\textbf{Forward process}}\label{method_forward_process}
Given user representation $\bm{u}_i$, we can set $\bm{u}_{i,0}=\bm{u}_i$ as the initial state. The forward process constitutes a Markov chain to gradually add Gaussian noise into $\bm{u}_{i,0}$ as follows:
\begin{equation}
\small
\begin{split}
q(\bm{u}_{i,t}|\bm{u}_{i,t-1}) = \mathcal{N}(\bm{u}_{i,t};\sqrt{1-\beta_t}\bm{u}_{i,t-1},\beta_t\textbf{I}),
\end{split}
\label{forward_process_1}
\end{equation}
where $t\in\{1,\dots,T\}$ denotes the diffusion step, $\beta_t\in(0,1)$ controls the forward noise scale added at each step $t$, and $\mathcal{N}$ denotes the Gaussian distribution. Thanks to the reparameterization trick~\cite{aevb}, $\bm{u}_{i,t}$ can be reparameterized from $\bm{u}_{i,0}$ as follows:
\begin{equation}
\small
\begin{split}
q(\bm{u}_{i,t}|\bm{u}_{i,0}) = \mathcal{N}(\bm{u}_{i,t};\sqrt{\bar{\alpha}_t}\bm{u}_{i,0},(1-\bar{\alpha}_t)\textbf{I}),
\end{split}
\label{forward_process_2}
\end{equation}
where $\alpha_t = 1-\beta_t$, $\bar{\alpha}_t = \prod_{t'=1}^{t}\alpha_{t'}$ and $\bm{\epsilon}\sim\mathcal{N}(0,\textbf{I})$, then we can obtain $\bm{u}_{i,t} = \sqrt{\bar{\alpha}_t}\bm{u}_{i,0} + \sqrt{1 - \bar{\alpha}_t}\bm{\epsilon}$. However, adding too much noise to user representation $\bm{u}_{i,0}$ would hurt user's personalized preference modelling. Therefore, we design an exponential noise schedule for $1-\bar{\alpha}_t$ as follows:
\begin{equation}
\small
\begin{split}
1-\bar{\alpha}_t=\eta\cdot\left[1-\text{exp}\left(-\frac{\alpha_{\text{min}}}{T}-\frac{S_t(\alpha_{\text{max}}-\alpha_{\text{min}})}{2T^2}\right)\right],\quad t\in\{1,\dots,T\},
\end{split}
\label{noise_scale}
\end{equation}
where $S_t$ represents the set of $T$ uniformly spaced values from 1 to $2T+1$, $\eta$ is a hyper-parameter ranging from 0 to 1 that controls the added noise scales, and two hyper-parameters $\alpha_{\text{min}}$ and $\alpha_{\text{max}}$ representing the lower and upper bounds of the added noise.

\subsubsection{\textbf{Reverse process}}
As the core phase of diffusion processes, the reverse process aims to iteratively denoise $\hat{\bm{u}}_{i,T}$\footnote{We set $\hat{\bm{u}}_{i,T}=\bm{u}_{i,T}$ for brevity.} for $T$ steps to ultimately reconstruct the user representation $\hat{\bm{u}}_{i,0}$. Notably, we incorporate the user's interaction history $\mathcal{H}_i^s$ in the source domain as the preference guidance signal, which is explicitly injected to user representation step by step to guide the reverse process, and ultimately generate personalized user representation $\hat{\bm{u}}_{i,0}$ in the target domain, which can be formulated as:
\begin{equation}
\small
\begin{split}
p_{\theta}(\hat{\bm{u}}_{i,t-1}|\hat{\bm{u}}_{i,t},\bm{h}_i^s) = \mathcal{N}(\hat{\bm{u}}_{i,t-1};\bm{\mu}_\theta(\hat{\bm{u}}_{i,t},\bm{h}_i^s,t),\textstyle\sum_\theta(\hat{\bm{u}}_{i,t},\bm{h}_i^s,t)),
\end{split}
\label{reverse}
\end{equation}
where we use a Multi-Layer Perceptron (MLP) as the architecture of $\bm{\mu}_\theta(\hat{\bm{u}}_{i,t},\bm{h}_i^s,t)$ that takes $\hat{\bm{u}}_{i,t}$, the preference guidance signal $\bm{h}_i^s$ and the diffusion step embedding of $t$ as inputs to generate $\hat{\bm{u}}_{i,0}$.

\subsection{Training and Inference}
In the training phase, we utilize overlapping users $u_i^o\in\mathcal{U}^o$ to perform the preference-guided diffusion processes.
In the inference phase, given a cold-start user $u_i^{s\setminus o}\in\mathcal{U}^{s\setminus o}$, our goal is to generate personalized user representations for cold-start users based on their interaction history in the source domain through reverse process.

\subsubsection{\textbf{Training}}
After performing diffusion processes on overlapping users and generating personalized user representation $\hat{\bm{u}}_{i,0}^o$ in the target domain, we choose the widely used inner product operation to implement the score function $\hat{y}_{ij}^t=\hat{\bm{u}}_{i,0}^o\bm{v}_j^t$ for user-item rating prediction. Following the previous CDR methods~\cite{emcdr,catn,ptupcdr,cdrnp}, we adopt the Mean-Square Error (MSE) loss for optimization:
\begin{equation}
\small
\begin{split}
\mathcal{L}_{rec}=\frac{1}{|\mathcal{Y}^o|}\sum_{y_{ij}^t\in\mathcal{Y}^o}(y_{ij}^t-\hat{y}_{ij}^t)^2,
\end{split}
\label{loss_L_rec}
\end{equation}
where $\mathcal{Y}^o=\{y_{ij}^t|u_i^o\in\mathcal{U}^o,v_j^t\in\mathcal{V}^t\}$ denotes the ratings of overlapping users in the target domain. Besides, we improve upon $\mathcal{L}_{t-1}$ in Eq.~(\ref{pre_loss}) with preference guidance signal $\bm{h}_i^s$ to force the prior distribution $p_\theta(\hat{\bm{u}}_{i,t-1}^o|\hat{\bm{u}}_{i,t}^o,\bm{h}_i^s)$ to approximate the tractable posterior distribution $q(\bm{u}_{i,t-1}^o|\bm{u}_{i,t}^o,\bm{u}_{i,0}^o)$ via KL divergence:
\begin{equation}
\small
\begin{split}
\mathcal{L}_{t-1}=D_{\text{KL}}(q(\bm{u}_{i,t-1}^o|\bm{u}_{i,t}^o,\bm{u}_{i,0}^o)\Vert p_\theta(\hat{\bm{u}}_{i,t-1}^o|\hat{\bm{u}}_{i,t}^o,\bm{h}_i^s)),
\end{split}
\label{loss_L_t-1_1}
\end{equation}

Using Bayes' theorem, the posterior distribution $q(\bm{u}_{i,t-1}^o|\bm{u}_{i,t}^o,\bm{u}_{i,0}^o)$ could be solved as the following closed form~\cite{udmaup}:
\begin{equation}
\small
\begin{split}
&q(\bm{u}_{i,t-1}^o|\bm{u}_{i,t}^o,\bm{u}_{i,0}^o)=\mathcal{N}(\bm{u}_{i,t-1}^o;\tilde{\bm\mu}_t(\bm{u}_{i,t}^o,\bm{u}_{i,0}^o),\tilde{\beta}_t\textbf{I}),\\
\text{where}\quad&\tilde{\bm\mu}_t(\bm{u}_{i,t}^o,\bm{u}_{i,0}^o)=\frac{\sqrt{\bar\alpha_{t-1}}\beta_t}{1-\bar\alpha_t}\bm{u}_{i,0}^o+\frac{\sqrt{\alpha_t}(1-\bar\alpha_{t-1})}{1-\bar\alpha_t}\bm{u}_{i,t}^o,\\
\text{and}\quad&\tilde{\beta}_t=\frac{1-\bar\alpha_{t-1}}{1-\bar\alpha_t}\beta_t,
\end{split}
\label{loss_forward}
\end{equation}
where $\tilde{\bm\mu}_t(\bm{u}_{i,t}^o,\bm{u}_{i,0}^o)$ and $\tilde{\beta}_t\textbf{I}$ represent the mean and variance of $q(\bm{u}_{i,t-1}^o|\bm{u}_{i,t}^o,\bm{u}_{i,0}^o)$. Following~\cite{ddpm}, we set $\textstyle\sum_\theta(\hat{\bm{u}}_{i,t}^o,\bm{h}_i^s,t)=\sigma_t^2\textbf{I}=\tilde{\beta}_t\textbf{I}$ in $p_{\theta}(\hat{\bm{u}}_{i,t-1}^o|\hat{\bm{u}}_{i,t}^o,\bm{h}_i^s)$ to match the variance of Eq.~(\ref{reverse}) and Eq.~(\ref{loss_forward}). Besides, we further reparameterizing $\bm{\mu}_\theta(\hat{\bm{u}}_{i,t}^o,\bm{h}_i^s,t)$ as follows:
\begin{equation}
\small
\begin{split}
\bm{\mu}_\theta(\hat{\bm{u}}_{i,t}^o,\bm{h}_i^s,t) = \frac{1}{\sqrt{\alpha_t}}\left(\hat{\bm{u}}_{i,t}^o-\frac{1-\alpha_t}{\sqrt{1-\bar{\alpha}_t}}\bm{\epsilon}_\theta(\hat{\bm{u}}_{i,t}^o,\bm{h}_i^s,t)\right),
\end{split}
\label{loss_mean}
\end{equation}

Therefore, the training objective of $\mathcal{L}_{t-1}$ in Eq.~(\ref{loss_L_t-1_1}) at step $t$ can be calculated by:
\begin{equation}
\small
\begin{split}
\mathcal{L}_{t-1}=\mathbb{E}_q\left[\frac{1}{2\sigma_t^2}\Vert\tilde{\bm\mu}_t(\bm{u}_{i,t}^o,\bm{u}_{i,0}^o)-\bm{\mu}_\theta(\hat{\bm{u}}_{i,t}^o,\bm{h}_i^s,t)\Vert^2\right]+\mathcal{C},
\end{split}
\label{loss_L_t-1_2}
\end{equation}
where $\mathcal{C}$ is a constant independent of the model parameter $\theta$. As suggested by\cite{ddpm}, Eq.~(\ref{loss_L_t-1_2}) can be further parameterized as:
\begin{equation}
\small
\begin{split}
\mathcal{L}_{t-1}=\mathbb{E}_{\bm{u}_{i,0}^o,\bm\epsilon}\left[\frac{\beta_t^2}{2\sigma_t^2\alpha_t(1-\bar\alpha_t)}\Vert\bm\epsilon-\bm\epsilon_\theta(\hat{\bm{u}}_{i,t}^o,\bm{h}_i^s,t)\Vert^2\right]+\mathcal{C},
\end{split}
\label{loss_L_t-1_3}
\end{equation}
In stead of predicting the added noise $\bm\epsilon$ in Eq.~(\ref{loss_mean})\footnote{Our model is optimized by predicting $\hat{\bm{u}}_{i,0}^o$ instead of $\bm\epsilon$ because: 1) as suggested by~\cite{dreamrec}, predicting $\hat{\bm{u}}_{i,0}^o$ and predicting $\bm\epsilon$ are equivalent due to $\bm{u}_{i,t}^o=\sqrt{\bar\alpha_t}\bm{u}_{i,0}^o+\sqrt{1-\bar\alpha_t}\bm\epsilon$; and 2) the key objective of CDR is to predict $\hat{\bm{u}}_{i,0}^o$ in the target domain to make recommendation, thus predicting $\hat{\bm{u}}_{i,0}^o$ is more appropriate for our CDR task.}, we employ another reparameterization that predicts user representation $\hat{\bm{u}}_{i,0}^o$:
\begin{equation}
\small
\begin{split}
\bm{\mu}_\theta(\hat{\bm{u}}_{i,t}^o,\bm{h}_i^s,t)=\sqrt{\bar\alpha_{t-1}}f_\theta(\hat{\bm{u}}_{i,t}^o,\bm{h}_i^s,t)+\frac{\sqrt{\alpha_t}(1-\bar\alpha_{t-1})}{\sqrt{1-\bar\alpha_t}}\bm\epsilon,
\end{split}
\label{}
\end{equation}
Therefore, Eq.~(\ref{loss_L_t-1_3}) converts to another format:
\begin{equation}
\small
\begin{split}
\mathcal{L}_{t-1}=\mathbb{E}_{\bm{u}_{i,0}^o,\bm\epsilon}\left[\frac{\bar\alpha_{t-1}}{2\sigma_t^2}\Vert\bm{u}_{i,0}^o-f_\theta(\hat{\bm{u}}_{i,t}^o,\bm{h}_i^s,t)\Vert^2\right]+\mathcal{C},
\end{split}
\label{loss_L_t-1_4}
\end{equation}

In summary, the overall optimization function $\mathcal{L}$ can be defined as follows:
\begin{equation}
\small
\begin{split}
\mathcal{L}=\mathcal{L}_{rec}+\lambda\mathcal{L}_{t-1},
\end{split}
\label{loss_L}
\end{equation}
where $\lambda$ is a hyper-parameter ranging from 0 to 1.

Note that DMCDR's training phase concentrates on reconstructing the personalized user representation in the target domain. Furthermore, in addition to the conditional diffusion model, guided diffusion also needs an unconditional diffusion model~\cite{dmbgans} to  
further ensure the generated user representation align closely with preference guidance signal.
Inspired by~\cite{cfdg}, we jointly train the conditional and unconditional diffusion models with a \textit{classifier-free guidance} strategy for simplicity. Specifically, in the training phase, we can randomly replace the preference guidance signal $\bm{h}_i^s$ by a null token $\emptyset$ with a probability hyper-parameter $p_{\text{uncond}}$ to train the unconditional diffusion model:
\begin{equation}
\small
\begin{split}
\textbf{mask}=&\frac{\textit{sign}(\bm{r}-p_{\text{uncond}})+1}{2},\quad \bm{r}\sim \text{Uniform}(0,1),\\
\bm{h}_i^s=&\bm{h}_i^s\odot \textbf{mask}+\emptyset\odot (1-\textbf{mask}),
\end{split}
\label{}
\end{equation}
where \textit{sign} is a judgment function that returns 1 if positive, otherwise -1. The training phase of DMCDR is shown in Algorithm~\ref{alg1}.

\begin{algorithm}[tb]
\caption{The training phase of DMCDR}
\label{alg1}
\textbf{Input}: Overlapping users $\mathcal{U}^o$; Items set $\mathcal{V}^s$, $\mathcal{V}^t$; Rating matrix $\mathcal{Y}^t$. \\
\textbf{Output}: Model parameters $\theta$.\quad\quad
\begin{algorithmic}[1] 
\STATE Initialize all model parameters. \\
\STATE \textbf{repeat} \\
\STATE\quad Generate $\bm{h}_i^s$ in Eq.~(\ref{encoder_1})-(\ref{encoder_2}). \\ 
\STATE\quad Perform unconditional training with $p_{\text{uncond}}:\bm{h}_i^s=\emptyset$.\\
\STATE\quad Sample $t\sim \text{Uniform}(\{1,\dots,T\})$.\\
\STATE\quad Sample $\epsilon\sim \mathcal{N}(0,\textbf{I})$.\\
\STATE\quad Perform $q(\bm{u}_{i,t}^o|\bm{u}_{i,0}^o)$ in Eq.~(\ref{forward_process_2}).\\
\STATE\quad Perform $p_{\theta}(\hat{\bm{u}}_{i,t-1}^o|\hat{\bm{u}}_{i,t}^o,\bm{h}_i^s)$ in Eq.~(\ref{reverse}).
\STATE\quad Calculate $\mathcal{L}_{rec}$ in Eq.~(\ref{loss_L_rec}) and $\mathcal{L}_{t-1}$ in Eq.~(\ref{loss_L_t-1_4}).
\STATE\quad Optimize $\theta$ with the overall loss $\mathcal{L}$ in Eq.~(\ref{loss_L}).
\STATE\textbf{until} converged
\end{algorithmic}
\end{algorithm}

\begin{algorithm}[tb]
\caption{The inference phase of DMCDR}
\label{alg2}
\textbf{Input}: Cold-start users $\mathcal{U}^{s\setminus o}$; Items set $\mathcal{V}^s$, $\mathcal{V}^t$; Model parameters $\theta$. \\
\textbf{Output}: Prediction rating $\hat{y}_{ij}^t$.\quad\quad
\begin{algorithmic}[1] 
\STATE Generate $\bm{h}_i^s$ in Eq.~(\ref{encoder_1})-(\ref{encoder_2}). \\
\STATE\textbf{for} $t=T,\dots,1$ \textbf{do}
\STATE\quad Sample $\bm{z}\sim\mathcal{N}(0,\textbf{I})$ if $t>1$, else $\bm{z}=0$.\\
\STATE\quad Control the strength of $\bm{h}_i^s$ in Eq.~(\ref{control_strength}).\\
\STATE\quad Perform $p_{\theta}(\bm{u}_{i,t-1}^{s\setminus o}|\bm{u}_{i,t}^{s\setminus o},\bm{h}_i^s)$ in Eq.~(\ref{inference_reverse}).\\
\STATE\textbf{end for}\\
\STATE\textbf{return} $\bm{u}_{i,0}^{s\setminus o}$
\end{algorithmic}
\end{algorithm}

\subsubsection{\textbf{Inference}}
In the inference phase, we aim to generate the personalized user representation $\bm{u}_{i,0}^{s\setminus o}$ for cold-start users with their interaction history $\mathcal{H}_i^s$ in source domain through reverse process.

Typically, previous recommendation methods~\cite{dreamrec,difashion,mcdrec} randomly sample a Gaussian noise for the reverse process (described in Section~\ref{pre_inference}), possibly guided by other conditions. However, pure noise lacks user's personalized preference, and directly performing the reverse process on it will lead to the generation of sub-optimal user representation. Consequently, we first initialize the cold-start users representation $\bm{u}_{i,T}^{s\setminus o}=\bm{u}_i^{s\setminus o}$ instead of sampling Gaussian noise, and then perform the reverse process guided by the preference guidance signal $\bm{h}_i^s$ for $T$ steps to generate the personalized user representation $\bm{u}_{i,0}^{s\setminus o}$ in the target domain.

Note that in the inference phase, we modify $f_\theta(\bm{u}_{i,t}^{s\setminus o},\bm{h}_i^s,t)$ to control the strength of the preference guidance signal $\bm{h}_i^s$ as:
\begin{equation}
\small
\begin{split}
\tilde{f}_\theta(\bm{u}_{i,t}^{s\setminus o},\bm{h}_i^s,t)=(1+\omega)f_\theta(\bm{u}_{i,t}^{s\setminus o},\bm{h}_i^s,t)-\omega f_\theta(\bm{u}_{i,t}^{s\setminus o},\emptyset,t),
\end{split}
\label{control_strength}
\end{equation}
where $\omega$ is a hyper-parameter to control the strength of $\bm{h}_i^s$.

Based on Eq.~(\ref{loss_forward}) and Eq.~(\ref{control_strength}), we can rewrite the reverse process for one step as follows:
\begin{equation}
\small
\begin{split}
\bm{u}_{i,t-1}^{s\setminus o}=\frac{\sqrt{\bar\alpha_{t-1}}\beta_t}{1-\bar\alpha_t}\tilde{f}_\theta(\bm{u}_{i,t}^{s\setminus o},\bm{h}_i^s,t)+\frac{\sqrt{\alpha_t}(1-\bar\alpha_{t-1})}{1-\bar\alpha_t}\bm{u}_{i,t}^{s\setminus o}+\sqrt{\tilde{\beta}_t}\bm{z},
\end{split}
\label{inference_reverse}
\end{equation}
where $\bm{z}\sim\mathcal{N}(0,\textbf{I})$.
After generating the personalized user representation $\bm{u}_{i,0}^{s\setminus o}$ in the target domain through reverse process (described in Eq.~(\ref{inference_reverse})) guided by the preference guidance signal $\bm{h}_i^s$ for $T$ steps, we implement the score function $\hat{y}_{ij}^t=\bm{u}_{i,0}^{s\setminus o}\bm{v}_j^t$ for rating prediction. The inference phase of DMCDR is demonstrated in Algorithm~\ref{alg2}.

\section{Experiments}
In this section, we conduct extensive experiments on three real-world CDR scenarios to answer the following research questions:
\begin{itemize}
    \item \textbf{RQ1:} How does our DMCDR perform compared to the state-of-the-art recommendation methods?
    \item \textbf{RQ2:} How do different variants of DMs perform in CDR?
    \item \textbf{RQ3:} How do different components of DMCDR benefit its recommendation performance?
    \item \textbf{RQ4:} How do the hyper-parameter settings of DMCDR impact the performance, such as the inference step $T'$, the preference guidance strength $\omega$, the noise scale $\eta$ and the length of $\mathcal{H}_i^s$? 
    \item \textbf{RQ5:} How does the computational cost of DMCDR?
\end{itemize}

\subsection{Experimental Settings}

\begin{table}[t]
\centering
\footnotesize
\caption{Statistics of three CDR scenarios (\textit{Overlap} denotes the number of overlapping users).}
\setlength{\tabcolsep}{4.5pt}
{
\begin{tabular*}{0.47 \textwidth}{@{\extracolsep{\fill}}@{}lccccc@{}}
\toprule
{\bf Scenarios}  & \bf Domain  &  {\bf Users}  &  {\bf Overlap}  &  {\bf Items}  &  {\bf Ratings}\\
\midrule
\multirow{2}{*}{\textit{\textbf{Movie}}$\to$\textit{\textbf{Music}}}  & Movie  &123,960   &\multirow{2}{*}{18,031}   & 
 {50,052}  &1,697,533\\ 
& Music  &75,258  &  &  {64,443}  &1,097,592\\
\midrule
\multirow{2}{*}{\textit{\textbf{Book}}$\to$\textit{\textbf{Movie}}}  & Book  &603,668   &\multirow{2}{*}{37,388}   &{367,982}  &8,898,041\\ 
&Movie   &123,960  &  &  {50,052}  &1,697,533\\ 
\midrule
\multirow{2}{*}{\textit{\textbf{Book}}$\to$\textit{\textbf{Music}}}  & Book  &603,668   &\multirow{2}{*}{16,738}   &{367,982}  &8,898,041\\ 
&Music   &75,258  &  &  {64,443}  &1,097,592\\ 
\bottomrule
\end{tabular*}
}
\label{dataset}
\end{table}

\subsubsection{\textbf{Datasets}}
Following previous works~\cite{sscdr,catn,tmcdr,ptupcdr,cdrnp}, we conduct comprehensive experiments on Amazon review dataset\footnote{\url{http://jmcauley.ucsd.edu/data/amazon/}}, which contains rating data from 0 to 5, reflecting user preference for particular items. 
We select three relevant domains, including Books (termed as \textit{Book}), Movie and TV (termed as \textit{Movie}), CDs and Vinyl (termed as \textit{Music}) to form three CDR scenarios as scenario 1: \textit{Movie}$\to$\textit{Music}, scenario 2: \textit{Book}$\to$\textit{Movie} and scenario 3 : \textit{Book}$\to$\textit{Music}. While many existing works~\cite{tslcdr,sscdr,jpedet} only evaluate a subset of data, we directly use all available data to simulate the real-world application. The detailed information of three CDR scenarios is shown in Table~\ref{dataset}.

\subsubsection{\textbf{Baselines}}
We compare DMCDR with the following competitive baselines: (1) \textbf{TGT}~\cite{BPR} represents collaborative filtering methods based on matrix factorization (MF), which only uses interactions in the target domain. (2) \textbf{CMF}~\cite{cmf} is a collective matrix factorization method, which shares user representations across the source and target domains. (3) \textbf{EMCDR}~\cite{emcdr} is the first mapping-based method, which trains matrix factorization for users and items in both domains and then utilizes a mapping function to transfer user preference. (4) \textbf{DCDCSR}~\cite{dcdcsr} employs the MF models to generate user and item latent factors and then employs the DNN to map the latent factors across domains or systems. (5) \textbf{CATN}~\cite{catn} develops a review-based model to match aspect-level correlations of user and item across domains. (6) \textbf{SSCDR}~\cite{sscdr} extends EMCDR by designing a semi-supervised manner for metric space mapping and multi-hop neighborhood inference. (7) \textbf{DiffCDR}~\cite{diffcdr} employs DMs to generate user representations in the target domain without transferring user's personalized preference. (8) \textbf{PTUPCDR}~\cite{ptupcdr} follows the meta-learning paradigm with personalized mapping functions for each user to transfer their personalized preference. (9) \textbf{REMIT}~\cite{remit} further extends PTUPCDR to capture user's multiple interests with multiple mapping functions. (10) \textbf{CDRNP}~\cite{cdrnp} is the state-of-the-art method for CDR which follows the meta-learning paradigm and adopts neural process to capture the preference correlations among the overlapping and cold-start users.

\begin{table*}[t]
\footnotesize
\centering
\caption{Performance comparison of CDR to cold-start users in three CDR scenarios. The best results are highlighted in bold, while the second-best results are underlined. \% Improve represents the relative improvements of DMCDR over the best baseline.}
\label{mainexperiment}
\setlength\tabcolsep{2.0pt}
\begin{tabular*}{1 \textwidth}
{@{\extracolsep{\fill}}@{}lcccccc|cccccc|cccccc@{}}
\toprule
&
\multicolumn{6}{c}{\textit{\textbf{Movie}}$\to$\textit{\textbf{Music}}} & \multicolumn{6}{c}{\textit{\textbf{Book}}$\to$\textit{\textbf{Movie}}} & \multicolumn{6}{c}{\textit{\textbf{Book}}$\to$\textit{\textbf{Music}}}    \\
\cmidrule(r){2-7}\cmidrule(r){8-13}\cmidrule(r){14-19} \bf Methods &
\multicolumn{2}{c}{ \bf{20\%}} & \multicolumn{2}{c}{ \bf{50\%}} & \multicolumn{2}{c}{ \bf{80\%}} &  \multicolumn{2}{c}{ \bf{20\%}} & \multicolumn{2}{c}{ \bf{50\%}} & \multicolumn{2}{c}{ \bf{80\%}} &  \multicolumn{2}{c}{ \bf{20\%}} & \multicolumn{2}{c}{ \bf{50\%}} & \multicolumn{2}{c}{ \bf{80\%}} \\
\cmidrule(r){2-3}\cmidrule(r){4-5}\cmidrule(r){6-7}
\cmidrule(r){8-9}\cmidrule(r){10-11}\cmidrule(r){12-13}
\cmidrule(r){14-15}\cmidrule(r){16-17}\cmidrule(r){18-19}&
\bf{MAE} & \bf{RMSE} & \bf{MAE} & \bf{RMSE} & \bf{MAE} & \bf{RMSE} & \bf{MAE} & \bf{RMSE} & \bf{MAE} & \bf{RMSE} & \bf{MAE} & \bf{RMSE} & \bf{MAE} & \bf{RMSE} & \bf{MAE} & \bf{RMSE} & \bf{MAE} & \bf{RMSE} \\
\midrule
TGT  &   4.4803   &   5.1580   &   4.4989   &   5.1736   &   4.5020   &   5.1891   & 
 4.1831   &   4.7536   &   4.2288   &   4.7920   &   4.2123   &   4.8149   &   4.4873   &   5.1672   &   4.5073   &   5.1727   &   4.5204   &   5.2308  \\

CMF    &  1.5209 &   2.0158    &   1.6893    &    2.2271 &   2.4186  &   3.0936 & 
 1.3632   &   1.7918 &  1.5813   &    2.0886  &    2.1577    &    2.6777  &  1.8284  &  2.3829  &  2.1282  &  2.7275  &  3.0130  &  3.6948  \\
DCDCSR &  1.4918 &   1.9210    &   1.8144    &    2.3439 &   2.7194  &   3.3065 & 
 1.3971   &   1.7346 &  1.6731   &    2.0551  &    2.3618    &    2.7702  &  1.8411  &  2.2955  &  2.1736  &  2.6771  &  3.1405  &  3.5842  \\
SSCDR   &    1.3017 &   1.6579    &   1.3762    &    1.7477 &   1.5046  &   1.9229 & 
 1.2390   &   1.6526 &  1.2137   &    1.5602  &    1.3172    &    1.7024  &  1.5414  &  1.9283  &  1.4739  &  1.8441  &   1.6414  &  2.1403  \\
CATN  &  1.2671 &   1.6468 &  1.4890 &  1.9205 &  1.8182  &  2.2991  &  1.1249 &  1.4548   &  1.1598 &  1.4826 &  1.2672 &  1.6280  &  1.3924  &  1.7399  &  1.6023  &  2.0665  &  1.9571  &  2.5623 \\
EMCDR   &  1.2350 &   1.5515 &  1.3277 &  1.6644 &  1.5008  &  1.8771  &
 1.1162 &  1.4120   &  1.1832 &  1.4981 &  1.3156 &  1.6433  &  1.3524  &  1.6737  &  1.4723  &  1.8000  &  1.7191  &  2.1119  \\
PTUPCDR  &   1.1504 &   1.5195 &  1.2804 &  1.6380 &  1.4049  &  1.8234  &   0.9970 &  1.3317   &  1.0894 &  1.4395 &  1.1999 &  1.5916  &   1.2286  &  1.6085  &  1.3764  &  1.7447  &  1.5784  &  2.0510 \\
DiffCDR &   1.0435 &   1.3840 &  1.2367 &  1.6859 &  1.5606  &  2.1754  &   0.9476 &  1.2338   &  0.9953 &  1.3155 &  1.0846 &  1.4695  &   1.1220  &  1.5390  &  1.3077  &  1.8255  &  1.5871  &  2.2110 \\
REMIT &  0.9393  &  1.2709   &   1.0437   &  1.4580  &  1.2181  &  1.6601  &   \underline{0.8759} &  1.1650   &  0.9172 &  1.2379 &  1.0055 &  1.3772  &  1.3749  &  1.9940  &  1.4401  &  2.0495  &  1.6396  &  2.2653 \\
CDRNP  &  \underline{0.7974} &  \underline{1.0638}   &  \underline{0.7969} &  \underline{1.0589} &  \underline{0.8280} &  \underline{1.0758}  &  0.8846 &  \underline{1.1327}   &  \underline{0.8946} &  \underline{1.1450} &  \underline{0.8970} &  \underline{1.1576}  &  \underline{0.7453}  &  \underline{0.9914}  &  \underline{0.7629}  &  \underline{1.0111}  &  \underline{0.7787}  &  \underline{1.0373} \\
\midrule
DMCDR &  \textbf{0.6173}  &  \textbf{0.9425}   &   \textbf{0.6015}   &  \textbf{0.9559}  &  \textbf{0.6450}  &  \textbf{0.9706}  &  \textbf{0.5904}  &  \textbf{0.9929}  &  \textbf{0.6050}  &  \textbf{0.9859}  &  \textbf{0.6227}  &  \textbf{1.0002}  &  \textbf{0.5360}  &  \textbf{0.8726}  &  \textbf{0.5563}  &  \textbf{0.8842}  &  \textbf{0.5589}  &  \textbf{0.8992} \\
{\%} Improve & 22.59{\%}  &  11.40{\%}  &  24.52{\%}  &  9.73{\%}  &  22.10{\%}  &  9.78{\%} & 32.60{\%}  &  12.34{\%}  &  32.37{\%}  &  13.90{\%}  &  30.58{\%}  &  13.60{\%}  & 28.08{\%}  &  11.98{\%}  &  27.08{\%}  &  12.55{\%}  &  28.23{\%}  &  13.31{\%}\\
\bottomrule
\end{tabular*}
\end{table*}


\subsubsection{\textbf{Evaluation Metrics}}
Following previous works~\cite{emcdr,ptupcdr,remit,cdrnp}, we use Mean Absolute Error (MAE) and Root Mean Square Error (RMSE) as evaluation metrics for performance comparison.

\subsubsection{\textbf{Implementation Details}}
We implement all models with Python 3.6 and PyTorch 1.9.0 on Tesla T4 GPU. For a fair comparison, we tune the hyper-parameters of each model following their original literature. For our DMCDR, the dimension of user/item representations is set to 64, the hidden size is set to 64, the batch size is set to 128, the learning rate is set to 0.01 and the unconditional training probability $p_{\text{uncond}}$ is set to 0.1 suggested by~\cite{cfdg,dreamrec}. The number of Transformer layer and MLP layer is chosen from $\left\{{2, 3, 4, 5, 6}\right\}$, and the length of $\mathcal{H}_i^s$ is chosen from $\left\{{10, 20, 30, 40, 50}\right\}$. The diffusion step $T$ is tuned in $\left\{{50, 100, 200, 500, 1000}\right\}$, and the preference guidance strength $\omega$ is chosen from $\left\{{0, 1, 2, 3, 4, 5}\right\}$. Besides, the noise lower bound $\alpha_{\text{min}}$ and upper bound $\alpha_{\text{max}}$ are set to 0.1 and 10, the noise scale $\eta$ in Eq.~(\ref{noise_scale}) is searched in $\left\{{0.1, 0.3, 0.5, 0.7, 0.9}\right\}$ and we fix $\lambda$ in Eq.~(\ref{loss_L}) at $1e-2$. All models are trained for 10 epochs to achieve convergence and the best hyper-parameters are tuned by grid search according to MAE. All experimental results are averaged over five random runs. Note that we take the sequential timestamps into account to avoid information leakage.

Following previous works~\cite{ptupcdr,remit,cdrnp}, we randomly remove all the ratings of a fraction of overlapping users in the target domain and regard them as test users (cold-start users), while the samples of other overlapping users are used to train our model. In our experiments, we set the proportions of test users (cold-start users) $\alpha$ as 20\%, 50\% and 80\% of the total overlapping users, respectively.

\subsection{Recommendation Performance (RQ1)}
In this section, we compare our DMCDR against baseline models on three CDR scenarios in Table~\ref{mainexperiment}, from which we can summarize the following observations.

Single domain recommendation model (\textit{e.g.}, TGT) cannot provide satisfactory results since it does not leverage auxiliary information in the source domain. This observation indicates the importance of transferring user preference across domains. 
Moreover, we also observe that DMCDR can effectively improve the recommendation performance compared with the mapping-based model (\textit{e.g.}, EMCDR) and meta-based model (\textit{e.g.}, CDRNP). Such improvements suggest that DMCDR is capable of integrating user preference by 
fully harnessing the information injection capability of DMs to explicitly inject preference into user representation modeling.
Another observation is that even with limited trainable overlapping users (\textit{e.g.}, $\alpha=80\%$), our proposed DMCDR can also achieve great recommendation performance. This indicates that our preference-guided diffusion model can be suitable for CDR scenarios with limited trainable data.
Overall, DMCDR significantly and consistently outperforms all baseline models, which demonstrates the superiority of our proposed preference-guided diffusion model in solving the cold-start problem for CDR.

\begin{figure}[t!]
\begin{center}
\includegraphics[width=8.5cm]{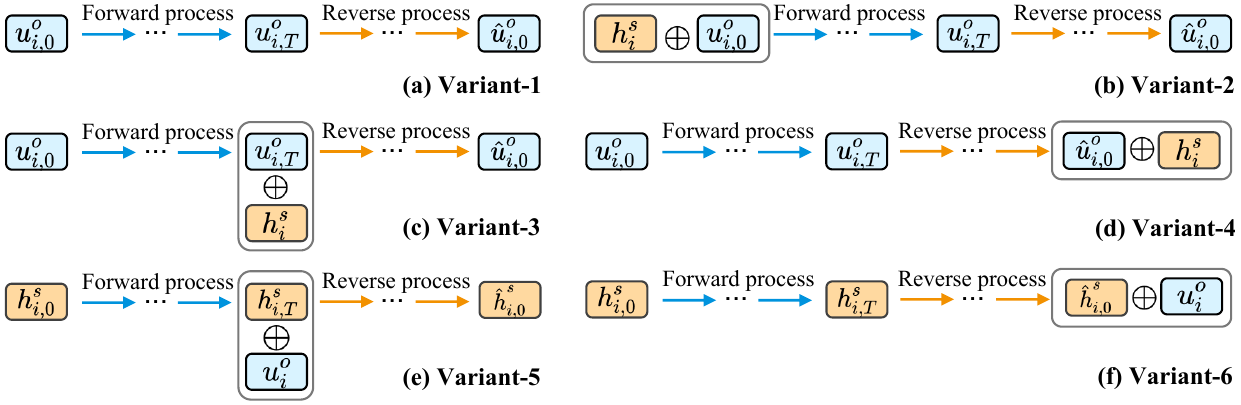}
\caption{An illustration of six DMs-based variants in the training phase, where $\oplus$ denotes the concatenation operation.}
\label{variant_figure}
\end{center}
\end{figure}

\begin{table}[t]
\centering
\footnotesize
\caption{Variant Study on three CDR scenarios.}
\label{variant}
\setlength\tabcolsep{2pt}{
\begin{tabular*}{0.47 \textwidth}{@{\extracolsep{\fill}}@{}lcccccc@{}}
\toprule
\multirow{2.5}{*}{\bf Variants}  &  
\multicolumn{2}{c}{\textit{\textbf{Movie}}$\to$\textit{\textbf{Music}}} & \multicolumn{2}{c}{\textit{\textbf{Book}}$\to$\textit{\textbf{Movie}}} &
\multicolumn{2}{c}{\textit{\textbf{Book}}$\to$\textit{\textbf{Music}}}\\ 
\cmidrule(r){2-3}\cmidrule(r){4-5}\cmidrule(r){6-7}&
\bf{MAE} & \bf{RMSE} & \bf{MAE} & \bf{RMSE} & \bf{MAE} & \bf{RMSE} \\
\midrule
Variant-1 &  0.6889   &  1.0103 &   0.7041 & 1.0671 & 0.6189 & 0.9578 \\
Variant-2  & 0.6641 & 0.9929 & 0.6717 & 1.0339 & 0.5809 & 0.9334  \\
Variant-3  & 0.6532 & 0.9707 & 0.6634 & 1.0375 & 0.5891 & 0.9234  \\
Variant-4  & 0.6756 & 0.9968 & 0.6855 & 1.0283 & 0.6067 & 0.9470  \\
Variant-5  & 0.6458 & 0.9833 & 0.6461 & 1.0305 & 0.5759 & 0.9480  \\
Variant-6  & 0.6619 & 1.0035 & 0.6736 & 1.0530 & 0.5837 & 0.9285  \\
DMCDR &   \textbf{0.6173} & \textbf{0.9425}  &  \textbf{0.5904} & \textbf{0.9929} & \textbf{0.5360} & \textbf{0.8726} \\

\bottomrule
\end{tabular*}
}
\end{table}

\subsection{Variant Study (RQ2)}\label{variant_study}
In order to explore the impact of different DMs-based variants on CDR performance, we conduct the variant study on six variant models (as shown in Figure~\ref{variant_figure}) with $\alpha=20\%$. For a fair comparison, we use the same operation as in Section~\ref{guidance_signal} to obtain the user preference representation $\bm{h}_i^s$ in the source domain for all variants:
\begin{itemize}
    \item \textbf{Variant-1:} This variant removes the guidance from $\bm{h}_i^s$, which only performs diffusion processes to $\bm{u}_{i,0}$.
    \item \textbf{Variant-2:} This variant concatenate $\bm{u}_{i,0}$ and $\bm{h}_i^s$ as the initial state of the forward process, then perform the diffusion processes.
    \item \textbf{Variant-3:} We concatenate $\bm{u}_{i,T}$ and $\bm{h}_i^s$ as the initial state $\hat{\bm{u}}_{i,T}$ of the reverse process, rather than the gradual guidance of $\bm{h}_i^s$. 
    \item \textbf{Variant-4:} We perform diffusion processes on $\bm{u}_{i,0}$ to obtain $\hat{\bm{u}}_{i,0}$, then concatenate $\hat{\bm{u}}_{i,0}$ and $\bm{h}_i^s$ to make recommendation.
    \item \textbf{Variant-5:} We perform forward process on $\bm{h}_{i,0}^s$\footnote{Similar to setting $\bm{u}_{i,0}=\bm{u}_i$ in Section~\ref{method_forward_process}, we set $\bm{h}_{i,0}^s=\bm{h}_i^s$.} to obtain $\bm{h}_{i,T}^s$, then concatenate $\bm{h}_{i,T}^s$ and $\bm{u}_i$ as $\hat{\bm{h}}_{i,T}^s$ to perform reverse process.
    \item \textbf{Variant-6:} We perform diffusion processes on $\bm{h}_{i,0}^s$ to obtain $\hat{\bm{h}}_{i,0}^s$, then concatenate $\hat{\bm{h}}_{i,0}^s$ and $\bm{u}_i$ to make recommendation.
\end{itemize}
where Variant-2, 5, 6 incorporate $\bm{h}_i^s$ into the forward process, which can be seen as transferring user preference in a denoising manner, removing irrelevant preference information in $\bm{h}_i^s$ through reverse process. From the results shown in Table~\ref{variant}, we have several interesting findings.
Compared to all six variants, our DMCDR achieves the best performance, which indicates that our designed preference-guided diffusion model is more suitable for handling cold-start problem in CDR. We also find that Variant-5 achieves the second best performance. Such observation suggests that DMs can remove irrelevant preference information for cold-start users in a denoising manner. It should be note that, as Variant-2, 5, 6 integrate $\bm{h}_i^s$ into the forward process, adding too much noise to $\bm{h}_i^s$ can damage the personalized information. Therefore, we should carefully control the added noise scales as discussed in Section~\ref{method_forward_process}.
In addition, Variant-1 has the worst performance, which highlights the necessity of transferring user preference across domains.

\begin{table}[t]
\centering
\footnotesize
\caption{Ablation Study on three CDR scenarios.}
\label{ablation}
\setlength\tabcolsep{2pt}{
\begin{tabular*}{0.47 \textwidth}{@{\extracolsep{\fill}}@{}lcccccc@{}}
\toprule
\multirow{2.5}{*}{\bf Variants}  &  
\multicolumn{2}{c}{\textit{\textbf{Movie}}$\to$\textit{\textbf{Music}}} & \multicolumn{2}{c}{\textit{\textbf{Book}}$\to$\textit{\textbf{Movie}}} &
\multicolumn{2}{c}{\textit{\textbf{Book}}$\to$\textit{\textbf{Music}}}\\ 
\cmidrule(r){2-3}\cmidrule(r){4-5}\cmidrule(r){6-7}&
\bf{MAE} & \bf{RMSE} & \bf{MAE} & \bf{RMSE} & \bf{MAE} & \bf{RMSE} \\
\midrule
\textit{w}/\textit{o} TF &   0.6438 &  0.9728 &  0.6286 & 1.0345 & 0.5573 & 0.9027 \\
\textit{w}/\textit{o} GS   &  0.6889   &  1.0103 &   0.7041 & 1.0671 & 0.6189 & 0.9578 \\
\textit{w}/\textit{o} DM &  0.8197   &  1.1094 &   0.8714 & 1.1476 & 0.7753 & 1.0223 \\
DMCDR &   \textbf{0.6173} & \textbf{0.9425}  &  \textbf{0.5904} & \textbf{0.9929} & \textbf{0.5360} & \textbf{0.8726} \\

\bottomrule
\end{tabular*}
}
\end{table}

\subsection{Ablation Study (RQ3)}
To verify the effectiveness of each component in DMCDR, we conduct the ablation study on three CDR scenarios with $\alpha=20\%$. Specifically, we develop three distinct model variants to establish a comparative analysis with our original model, which are outlined:
\begin{itemize}
    \item $\textit{w}/\textit{o}$ TF: This variant involves the removal of Transformer layer.
    \item $\textit{w}/\textit{o}$ GS: This variant excludes the preference guidance signal.
    \item $\textit{w}/\textit{o}$ DM: We remove our diffusion model and directly make recommendation with the concatenation of $\bm{u}_i$ and $\bm{h}_i^s$.
\end{itemize}
The ablation study results are shown in Table~\ref{ablation}. We can see that removal of Transformer layer leads to performance degradation. This finding validates the effectiveness of capturing user's personalized preference using the Transformer layer. Meanwhile, ablation of the preference guidance signal demonstrates its crucial role to guide the reverse process and improve the performance of our DMCDR. Among them, performance decreases the most after removing the diffusion model, indicating that our designed preference-guided diffusion model can 
fully exploit the information injection capability of DMs to explicitly inject preference into user representation, thus achieving the transfer of user preference across domains.

\begin{figure}[t]
\setlength{\abovecaptionskip}{0.cm}
	\begin{center}
        \subfigure
        {\begin{minipage}[b]{.49\linewidth}
        \centering
        \includegraphics[scale=0.29]{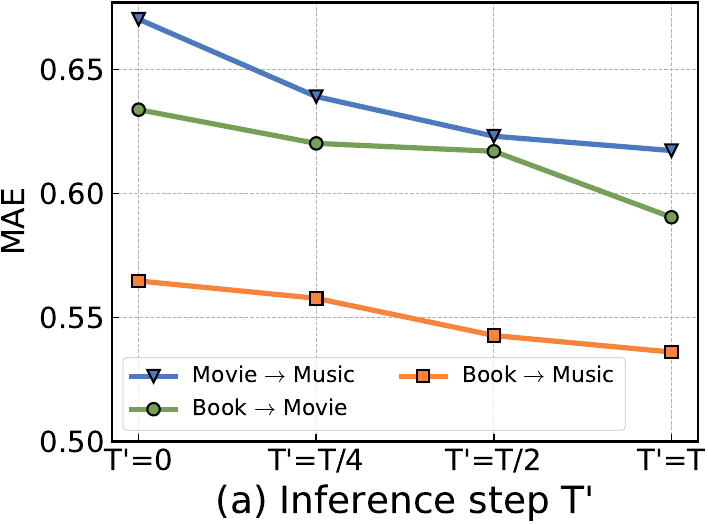}
        \end{minipage}}
        \subfigure
        {\begin{minipage}[b]{.49\linewidth}
        \centering
        \includegraphics[scale=0.29]{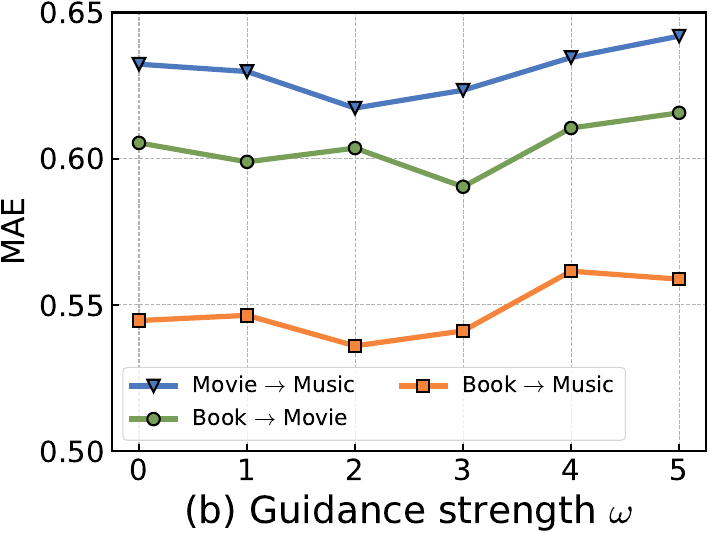}
        \end{minipage}}

	\caption{Effect of inference step $T'$ and guidance strength $\omega$.}
	\label{inference_guidance}
	\end{center}
\end{figure}

\begin{figure}[t]
\setlength{\abovecaptionskip}{0.cm}
	\begin{center}
        \subfigure
        {\begin{minipage}[b]{.49\linewidth}
        \centering
        \includegraphics[scale=0.29]{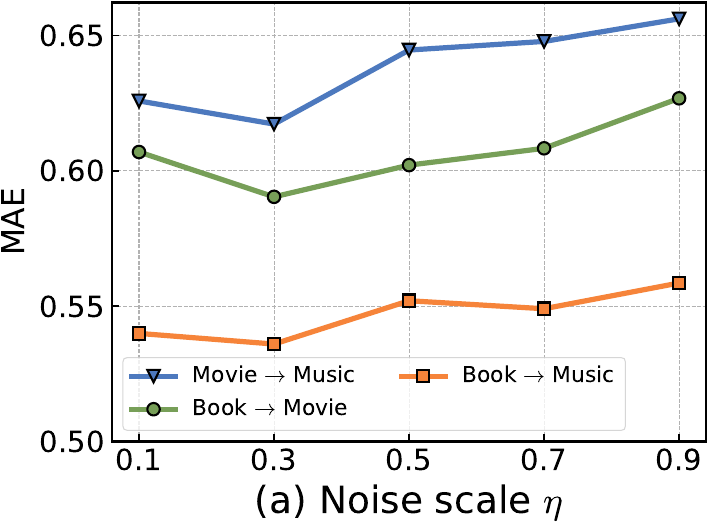}
        \end{minipage}}
        \subfigure
        {\begin{minipage}[b]{.49\linewidth}
        \centering
        \includegraphics[scale=0.29]{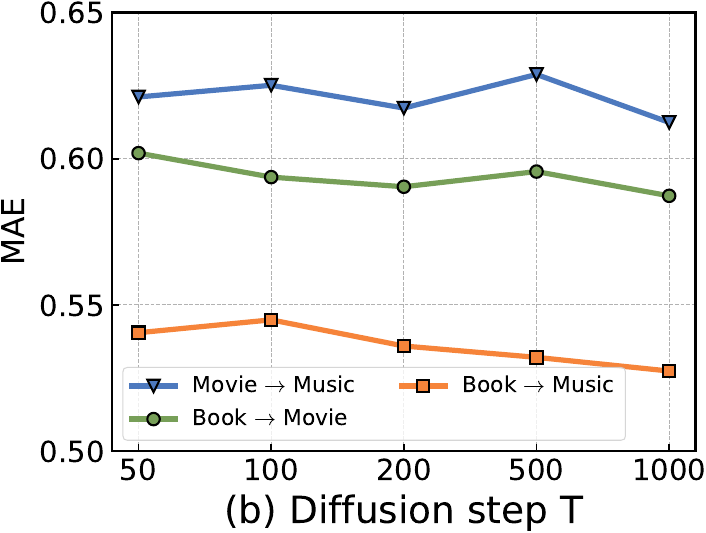}
        \end{minipage}}

	\caption{Effect of noise scale $\eta$ and diffusion step $T$.}
	\label{noise_diffusion}
	\end{center}
\end{figure}

\subsection{Hyper-parameter Analysis (RQ4)}
We further explore the effect of different hyper-parameter settings in DMCDR with $\alpha = 20\%$ in this section. 

\subsubsection{\textbf{Effect of inference step $T'$.}}
To verify the effectiveness of the preference guidance signal $\bm{h}_i^s$ to the reverse process, we vary the inference step $T'$ from 0 to $T$ during inference. From the results shown in Figure~\ref{inference_guidance}(a), we find that the performance of DMCDR improves as the inference step $T'$ increases, and achieve the best results when $T'=T$. This indicates that our designed preference-guided diffusion model can guide the reverse process step by step to generate the personalized user representation in the target domain. 

\subsubsection{\textbf{Effect of guidance strength $\omega$.}}
To achieve controllable guidance in DMCDR, we adopt the \textit{classifier-free guidance} strategy. And we investigate the sensitivity of DMCDR to the preference guidance strength $\omega$ in Eq.~(\ref{control_strength}). As the results shown in Figure~\ref{inference_guidance}(b), we find that increasing the value of $\omega$ initially leads to better performance. This observation indicates that a higher $\omega$ value can enhance the preference guidance to generate the personalized user representation in the target domain. However, as the value of $\omega$ continues to increase, we observe a decrease in performance, which indicates that too strong guidance strength may leads to the generation of low-quality user representation. Therefore, we should carefully adjust the value of guidance strength $\omega$.





\subsubsection{\textbf{Effect of noise scale $\eta$ and diffusion step $T$}}
In this section, we vary two hyper-parameters $\eta$ and $T$ to analyze their effects to DMCDR. From the results shown in Figure~\ref{noise_diffusion}(a), we find that the performance of DMCDR first rises as the nosie scale $\eta$ increases, indicating the effectiveness of denoising training with the guidance of user's interaction history during the reverse process. However, we find that the performance drops as the value of $\eta$ continues to increase. This observation shows that adding too much noise in the forward process will damage the personalized preference in user representation. Hence, we should choose a relatively small value of noise scale $\eta$. 
The results in Figure~\ref{noise_diffusion}(b) demonstrates that increasing the diffusion step $T$ has minimal impact on the performance due to the setting of a small value of $\eta$. Therefore, we choose $T=200$ to balance the performance and computational costs.

\subsubsection{\textbf{Effect of the length of $\mathcal{H}_i^s$}}
\begin{figure}[t]
\setlength{\abovecaptionskip}{0.cm}
	\begin{center}
        \subfigure
        {\begin{minipage}[b]{.32\linewidth}
        \centering
        \includegraphics[scale=0.23]{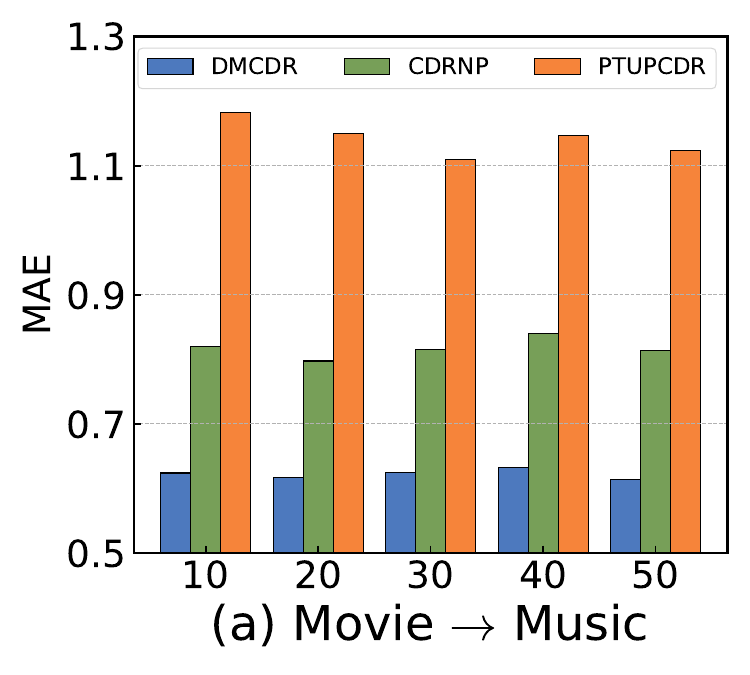}
        \end{minipage}}
        \subfigure
        {\begin{minipage}[b]{.32\linewidth}
        \centering
        \includegraphics[scale=0.23]{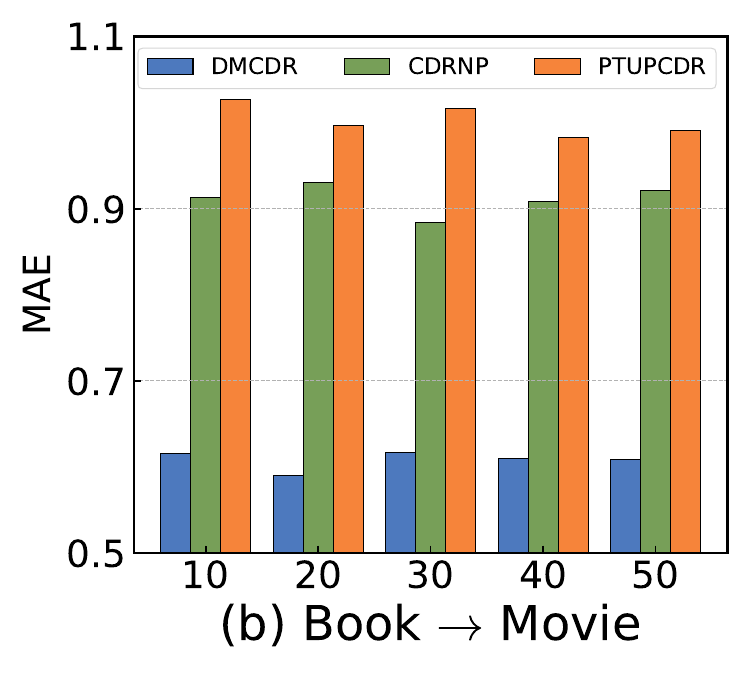}
        \end{minipage}}
        \subfigure
        {\begin{minipage}[b]{.32\linewidth}
        \centering
        \includegraphics[scale=0.23]{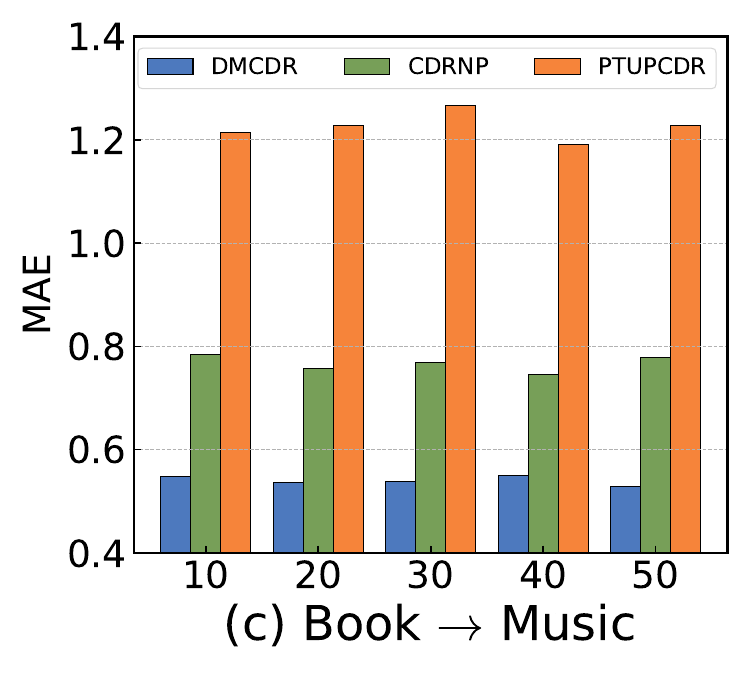}
        \end{minipage}}
        
	\caption{Effect of the length of $\mathcal{H}_i^s$.}
	\label{interaction_history}
	\end{center}
\end{figure}

To explore the effect of the length of user's interaction history $\mathcal{H}_i^s$ on model performance, we compare our DMCDR with the SOTA method CDRNP and a classical meta-learning method PTUPCDR. From the results shown in Figure~\ref{interaction_history}, we find that DMCDR consistently outperforms the other methods with different length of $\mathcal{H}_i^s$, indicating that our model can exhibits stable performance with different length of interaction history in the source domain. Another observation is that despite a short length of $\mathcal{H}_i^s$, our model still shows exceptional performance. The reason is that DMCDR can gradually generate personalized user representation in the target domain under the guidance of limited user's interaction history in the source domain through our designed preference-guided diffusion model.

\begin{figure}[t]
\setlength{\abovecaptionskip}{0.cm}
	\begin{center}
        \subfigure
        {\begin{minipage}[b]{.32\linewidth}
        \centering
        \includegraphics[scale=0.23]{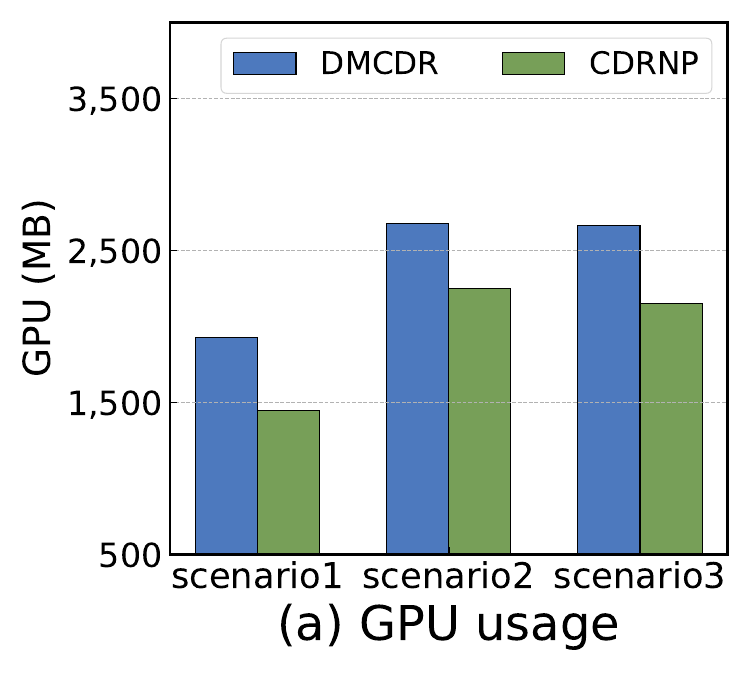}
        \end{minipage}}
        \subfigure
        {\begin{minipage}[b]{.32\linewidth}
        \centering
        \includegraphics[scale=0.23]{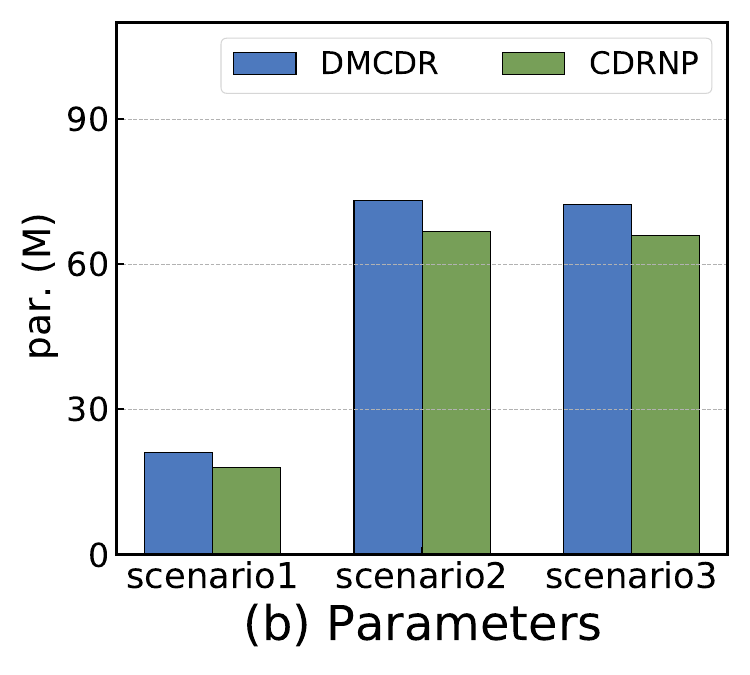}
        \end{minipage}}
        \subfigure
        {\begin{minipage}[b]{.32\linewidth}
        \centering
        \includegraphics[scale=0.23]{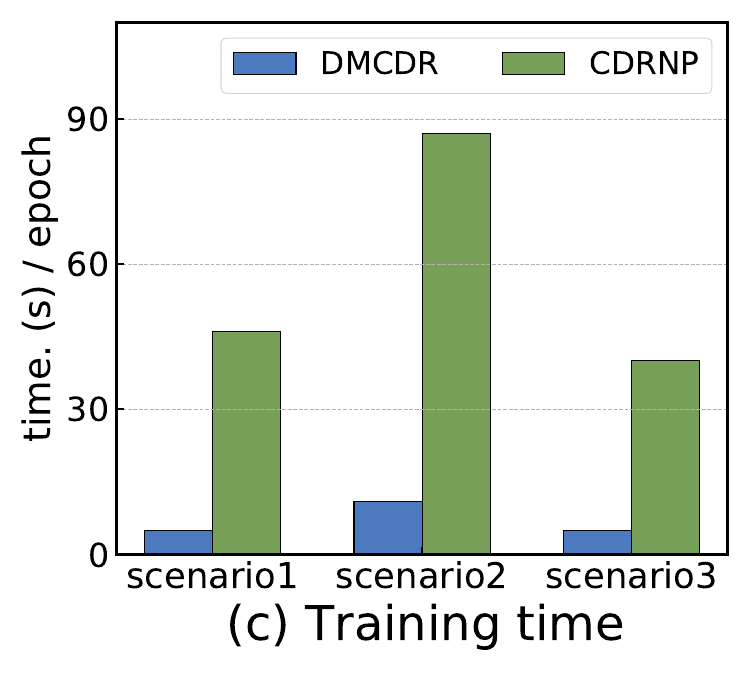}
        \end{minipage}}
        
	\caption{Computational cost on three CDR scenarios.}
	\label{computational_cost}
	\end{center}
\end{figure}

\subsection{Computational Cost Analysis (RQ5)}
We perform a comprehensive evaluation of the computational costs for our DMCDR and the SOTA method CDRNP, including GPU usage, parameters and training time for one epoch. From the results shown in Figure~\ref{computational_cost}, we observe that the running efficiency of DMCDR is not the best compared with CDRNP, but the efficiencies of DMCDR are still of the same order of magnitude.
In particular, despite the slightly higher GPU usage and parameters, the training time of DMCDR is much lower than that of CDRNP, which proves that our preference-guided diffusion model can save much training time from the computation. Besides, considering the significant performance advantage of our DMCDR, the computational costs are reasonable and acceptable.

\section{Related Work}
\subsection{Cross-Domain Recommendation}
Cross-domain recommendation (CDR)~\cite{emcdr,ptupcdr,cdrnp,conet,disencdr} aims to leverage the informative interaction history in the source domain to improve the recommendation performance in the target domain.  
Existing CDR methods are mainly divided into two types, one concentrates on tackling the data sparsity problem~\cite{conet,disencdr,kerkt,darec,tdar,cfaa} for overlapping users, the other focuses on alleviating the cold-start problem~\cite{emcdr,sscdr,tmcdr,cdrib,doml,ptupcdr,cdrnp} for cold-start (\textit{i.e.}, non-overlapping) users. In this paper, we focus on solving the cold-start problem.

Classical CDR methods aim to model cross-domain preference transfer through mapping-based paradigm~\cite{emcdr,sscdr,doml}. Specifically, EMCDR~\cite{emcdr} is the most popular method which first leverages the matrix factorization to learn the user/item embeddings, and then utilizes a mapping function to transfer user preference from one domain to the other. SSCDR~\cite{sscdr} further extends EMCDR by designing a semi-supervised manner to learn user/item mapping across domains. DOML~\cite{doml} develop a novel latent orthogonal mapping to extract user preference over multiple domains.
Recently, some CDR methods also follow the meta-based paradigm\cite{tmcdr,ptupcdr,remit,cdrnp}, which can still be regarded as a particular form of the mapping-based paradigm.  
Among them, TMCDR~\cite{tmcdr} and PTUPCDR~\cite{ptupcdr} propose meta networks for each user to transfer their personalized preference.
REMIT~\cite{remit} constructs a heterogeneous information network and employs different meta-path to get user's multiple interests. CDRNP~\cite{cdrnp} follows the meta-learning paradigm and leverages neural process to capture the preference correlations among the overlapping users and cold-start users.

\subsection{Diffusion Models in Recommendation}
Diffusion models (DMs)~\cite{dfsurvey} have shown remarkable generation capabilities for various generative tasks, including image synthesis~\cite{dmbgans,hrisld}, text generation~\cite{diffusion-lm}, and audio processing~\cite{wavegrad,diffwave}.  
Recently, some studies have demonstrated the explicit information injection capability of DMs in recommendation systems~\cite{drm,diff4rec,diffkg,pdrec}.
Specifically, DiffRec~\cite{drm} gradually generates personalized collaborative information in a denoising manner. DiffKG~\cite{diffkg} integrates the DMs with a data augmentation paradigm, enabling robust knowledge graph representation learning. Diff4Rec~\cite{diff4rec} applies DMs for data augmentation to capture the user-item relations. 
PDRec~\cite{pdrec} proposes three flexible plugin modules to jointly take full advantage of the diffusion-generating user preference on all items. 
Besides, recent advances consider adding some conditions over the generation process for conditional diffusion to realize the controllable generation~\cite{dreamrec,difashion,mcdrec}. DreamRec~\cite{dreamrec} uses DMs to generate the oracle items with the user's sequential behaviors and cast off negative sampling. MCDRec~\cite{mcdrec} tailors DMs with two technical modules to model the high-order multimodal knowledge. DiFashion~\cite{difashion} designs three conditions to guide the generation process to achieve the generation of multiple fashion images.

Different from prior CDR methods that focus on mapping features across domains, neglecting to explicitly model the preference integration process, we explore to leverage the information injection capability of DMs to explicitly inject preference into user representation, thus achieve the transfer of user preference and perform the new SOTA in CDR. We unleash the great potential of DMs to solve the cold-start problem in CDR.

\section{Conclusion and Future Work}
In this work, we propose a novel Preference-Guided Diffusion Model for CDR to cold-start users, termed as DMCDR.  
We apply DMs to generate personalized user representation in the target domain guided by user's interaction history in the source domain, thereby achieving the transfer of user preference across domains. Furthermore, we comprehensively explore the impact of six DMs-based variants on CDR. Extensive experiments demonstrate that DMCDR consistently outperforms SOTA methods and six DMs-based variants in three real-world CDR scenarios.

This work opens up a new research direction for CDR by employing DMs. In the future, we will continue to explore DMs in more challenging tasks, \textit{e.g.}, multi-domain recommendation.

\section*{Acknowledgement}
This work was funded by the National Key Research and Development Program of China (No.2021YFB3100600), the Youth Innovation Promotion Association of CAS (No.2021153), the National Natural Science Foundation of China (No.62406319), and the Postdoctoral Fellowship Program of CPSF (No.GZC20232968).

\balance
\bibliographystyle{ACM-Reference-Format}
\bibliography{sample-base-extend.bib}
\end{document}